\documentclass{moriond}

\usepackage{amsmath,amsfonts,amssymb}
\usepackage{colortbl}
\definecolor{Gray}{gray}{0.95}
\definecolor{RGray}{gray}{0.93}
\definecolor{CGray}{gray}{0.92}

\usepackage[table]{xcolor}
\usepackage[capitalise]{cleveref}
\usepackage{wrapfig}
\usepackage{booktabs}
\usepackage{lipsum}

\def\be{\begin{equation}}
\def\ee{\end{equation}}
\def\bea{\begin{eqnarray}}
\def\eea{\end{eqnarray}}

\newcommand{\Photo}{}

\begin{document}
\vspace*{4cm}
\title{LEPTON FLAVOUR UNIVERSALITY IN $\tau$ DECAYS}

\author{L. ALLWICHER and N. SELIMOVI\'C}

\address{Physik-Institut, Universit\"at Z\"urich\\
Winterthurerstrasse 190, 8057 Z\"urich, Switzerland}

\maketitle\abstracts{
The evidence for Lepton Flavour Universality (LFU) violation in semileptonic $B$-decays
has been rising over the past few years.
Relying on generic effective field theory (EFT) results, it has been shown that
models addressing the $B$-anomalies necessarily lead, at one-loop, to deviations
from LFU in $\tau$ decays at the few per-mil level.
Once a (renormalizable) UV model is specified, the leading-log EFT result receives
finite corrections from the matching at the UV scale.
We discuss such corrections in a motivated class of models for the B-anomalies, based on an extended
$SU(4) \times SU(3) \times SU(2) \times U(1)$ gauge sector.
In this scenario, we obtain precise predictions for the effective $W$-boson and $Z$-boson
couplings to leptons in terms of the masses and couplings of the new heavy fields.
We confirm a few per-mil deviation from universality,
within reach of future high-precision experiments.
}

\section{Introduction}


Lepton Flavour Universality tests provide stringent constraints on physics beyond the Standard Model (SM). In addition to the well-known $B$ anomalies, one can also investigate LFU violation in other observables, such as leptonic decays. These could then in turn be interpreted as a deviation from universality of the $W$ boson couplings to leptons.
The connection between the anomaly in $b\to c\tau\nu$ transitions and leptonic $\tau$ decays has been noted for the first time in~\cite{Feruglio:2016gvd,Feruglio:2017rjo}.
Basing on EFT arguments only, the authors have shown that, even in a New Physics (NP) model where $\tau$ decays are not affected at tree-level, they necessarily receive a modification at one-loop level which is of the order of a few per-mil.
Although this still lies well within the experimental precision we have today, it is worth studying the finite corrections to the leading-log EFT expression, in order to provide a precise prediction that may be tested in future experiments.
In this study we look at the leptonic LFU ratios~\cite{Pich:2013lsa}
\begin{align}
    \left|g^{(\tau)}_e/g^{(\mu)}_e\right|^2 &\equiv  \frac{\Gamma(\tau\to e\nu\bar{\nu})}{\Gamma(\mu\to e\nu\bar{\nu})}\left[\frac{\Gamma_{\rm SM}(\tau\to e\nu\bar{\nu})}{\Gamma_{\rm SM}(\mu\to e\nu\bar{\nu})}\right]^{-1},
\label{eq:one}
\end{align}
and the analogously defined $|g^{(\tau)}_\mu/g^{(\mu)}_e|^2$ and $|g^{(\tau)}_\mu/g^{(\tau)}_e|^2$, and compute them in a class of UV models aimed at explaining the $B$ anomalies, while addressing also the flavor puzzle of the SM.

\section{EFT description of leptonic decays}

The typical scale of leptonic decays of $\tau$ leptons or muons lies well below the EW scale, where both the SM and the NP contributions to the processes are best expressed in the so-called Low-Energy Effective Theory (LEFT).
This is the theory obtained when integrating out the heavy degrees of freedom of the SM, i.e. $W^\pm$, $Z$, $h$ and $t$, as well as hypothetical new heavy states.
It consists, in general, of all operators invariant under $SU(3)_{\mathrm{c}} \times U(1)_{\rm{Q}}$ constructed out of the light SM fields.
Truncating the expansion at dimension-six, the general Lagrangian can be written as
\be
\mathcal{L}_{\rm LEFT} = -\frac{2}{v^2} \sum_k \mathcal{C}_k \mathcal{O}_k \,.
\ee
In order to describe decays of the type $\ell_\beta \to \ell_\alpha \nu\bar\nu$, however, we only need the following operator,
\be
\mathcal{O}_{\nu e}^{V,LL} = (\bar\nu_L^\alpha \gamma_\mu \nu_L^\beta)(\bar e_L^\gamma \gamma^\mu e_L^\delta) \,,
\ee
where $\alpha$, $\beta$, $\gamma$, $\delta$ are flavour indices\,\footnote{Notice that we neglect the operator with a right-handed charged-lepton current since it is zero in the SM.}.
Under the assumption of small NP corrections, we can define
\be
R_{\beta\alpha} = \frac{\Gamma(\ell_\beta \to \ell_\alpha \nu \bar\nu)}{\Gamma_{\rm SM}(\ell_\beta \to \ell_\alpha \nu \bar\nu)} \equiv 1 + \delta R_{\beta\alpha} \,,
\ee
where
\be
\delta R_{\beta\alpha} \approx 2 \mbox{Re}[\mathcal{C}_{\nu e}^{V,LL}]^{\rm NP}_{\alpha\beta\beta\alpha} \,,
\ee
and we have used the fact that $[\mathcal{C}_{\nu e}^{V,LL}]^{\rm SM}_{\alpha\beta\beta\alpha} = 1$ in our conventions.
Since this operator is not subject to RGE running, we need to evaluate the coefficient at the electroweak scale in order to find the contribution we are looking for. In order to do so, the explicit NP model needs to be matched onto the Standard Model Effective Theory (SMEFT), an effective theory with the full SM field content and $SU(3)_\mathrm{c} \times SU(2)_\mathrm{L}\times U(1)_\mathrm{Y}$ gauge invariance, at the UV scale, and run down to the electroweak scale.
The SMEFT Lagrangian is normalised as
\begin{align}
\mathcal{L}_{\rm SMEFT}=-\frac{2}{v^2}\sum_k\, C_k O_k\,,
\end{align}
and at one-loop, the SMEFT-LEFT matching reads
\begin{align}
    [\mathcal{C}_{\nu e}^{V,LL}]^{\rm NP-full}_{\alpha\beta\beta\alpha}  &=  - 2 \sum_{\gamma=\alpha,\beta}
[C_{H\ell}^{(3)}]_{\gamma\gamma}(\mu) 
   + [C_{\ell\ell}]_{\alpha\beta\beta\alpha} + [C_{\ell\ell}]_{\beta\alpha\alpha\beta}\nonumber\\
 &-  \frac{ m_t^2 N_{\rm c}}{ 8\pi^2 v^2} \sum_{\gamma=\alpha,\beta}
[C_{\ell q}^{(3)}]_{\gamma\gamma 33} \left(1+2\log\frac{\mu^2}{m_t^2}\right)~, 
\end{align}
where $C_{H\ell}^{3}$, $C_{\ell\ell}$, and $C_{\ell q}^{(3)}$ are the coefficients of the operators
\begin{align}
    [O_{H\ell}^{(3)}]_{\alpha\beta} &= (\bar{\ell}^\alpha\gamma_\mu \sigma^I \ell^\beta) (H^\dagger i \overleftrightarrow{D^\mu}\sigma^I H)\,,  \\
    [O_{\ell\ell}]_{\alpha \beta \gamma\delta } &= (\bar \ell_{L}^{\alpha}   \gamma_{\mu}  \ell_{L}^{\beta})
    (\bar \ell_{L}^{\gamma}   \gamma^{\mu}  \ell_{L}^{\delta})\,, \\
    [O_{\ell q}^{(3)}]_{\alpha \beta i j } &= (\bar \ell_{L}^{\alpha}  \sigma^I \gamma_{\mu}  \ell_{L}^{\beta})
(\bar q_{L}^{i}  \sigma^I  \gamma^{\mu}  q_{L}^{j} )\,,
\end{align}
in the Warsaw basis~\cite{Grzadkowski:2010es}, which we will obtain by the one-loop matching of the NP onto the SMEFT. This way, we capture the important finite terms, in addition to the leading-log correction originally computed in \cite{Feruglio:2016gvd,Feruglio:2017rjo}, which is model independent and fixed by the RG running of the semileptonic operator $O_{\ell q}^{(3)}$.

\section{UV model}

It has been shown that the $U_1$ vector leptoquark (LQ) provides a good combined solution for both charged and neutral current $B$ anomalies, provided it has the approximately $U(2)$-like flavour structure in the couplings, with dominant couplings to the third generation of fermions~\cite{Cornella:2021sby}. Such a structure can be obtained naturally in the so-called 4321 models, based on an extended $SU(4) \times SU(3)' \times SU(2)_L \times U(1)_X$ gauge symmetry.
These provide a UV completion for the $U_1$, and the non-universality of the couplings is obtained by charging only the third fermion family under $SU(4)$, while the light generations are coupled to $SU(3)'$ (see Table \ref{tab:fieldcontent}).
\begin{wraptable}{r}{0.5\textwidth}
\hspace{0.1cm}
\begin{tabular}{|c|c|c|c|c|}
\hline
Field & $SU(4)$ & $SU(3)'$ & $SU(2)_L$ & $U(1)_{X}$ \\
\hline
\hline
$q^i_L$ & $\mathbf{1}$ & $\mathbf{3}$ & $\mathbf{2}$ & $1/6$ \\
$u^i_R$ & $\mathbf{1}$ & $\mathbf{3}$ & $\mathbf{1}$ & $2/3$  \\
$d^i_R$ & $\mathbf{1}$ & $\mathbf{3}$ & $\mathbf{1}$ & $-1/3$  \\
$\ell^i_L$ & $\mathbf{1}$ & $\mathbf{1}$ & $\mathbf{2}$ & $-1/2$ \\
$e^i_R$ & $\mathbf{1}$ & $\mathbf{1}$ & $\mathbf{1}$ & $-1$ \\ 
$\psi_L$ & $\mathbf{4}$ & $\mathbf{1}$ & $\mathbf{2}$ & $0$ \\ 
$\psi_R^{\pm}$ & $\mathbf{4}$ & $\mathbf{1}$ & $\mathbf{1}$ & $\pm1/2$ \\  \rowcolor{RGray}
$\chi_{L,R}$ & $\mathbf{4}$ & $\mathbf{1}$ & $\mathbf{2}$ & 0  \\  
\hline
\hline
$H$ & $\mathbf{1}$ & $\mathbf{1}$ & $\mathbf{2}$ & 1/2  \\     \rowcolor{RGray}
$\Omega_1$ & $\mathbf{\bar 4}$ & $\mathbf{1}$ & $\mathbf{1}$ & $-1/2$  \\ \rowcolor{RGray}
$\Omega_3$ & $\mathbf{\bar 4}$ & $\mathbf{3}$ & $\mathbf{1}$ & $1/6$  \\
\hline
\end{tabular}
\caption{Minimal field content of the model. Fields added to the SM matter content are shown in grey.}
\label{tab:fieldcontent}
\end{wraptable}
The SM gauge group is the subgroup of the 4321, with $SU(3)_c\times U(1)_Y\equiv[SU(4)\times SU(3)^\prime\times U(1)_X]_{\rm diag}$, and $SU(2)_L$ being the SM one. The hypercharge is defined in terms of the $U(1)_X$ charge, and the $SU(4)$ generator $T_4^{15}=\frac{1}{2\sqrt{6}}\mathrm{diag}(1,1,1,-3)$ by $Y=X+\sqrt{2/3}\,T_4^{15}$.

\noindent
In order to make the LQ couple also to the second generation, one needs to introduce vector-like (VL) fermions. These mix with one of the light generations, which can be chosen without loss of generality to be the second, through terms of the form
\be
    \mathcal{L} \supset \bar q_L^2 \Omega_3 \chi_R + \bar \ell_L^2 \Omega_1 \chi_R + \mbox{h.c.} \,,
\ee
which after the 4321 $\to$ SM breaking give rise to the mixing. Other realizations of the vector-like states and the mass mixing are possible. However, in the broken phase, all possibilities can be summarized with the generic mass term~\cite{Fuentes-Martin:2020hvc}
\begin{align}\label{eq:MassMix}
\mathcal{L}_{\rm mass}  = 
\bar\Psi_L^{q\,\prime}\, M_q\,Q_R+\bar\Psi_L^{\ell\,\prime}\, M_\ell\,L_R + \mbox{h.c.}\,,
\end{align}
with the left-handed fermions arranged as
\begin{align}\label{eq:Psiprime}
\Psi^{q\,\prime}_L=
\begin{pmatrix}
q_L^{\prime\, 2} &
q_L^{\prime\, 3} &
Q_L^\prime
\end{pmatrix}^\intercal
,\qquad
\Psi^{\ell\,\prime}_L=
\begin{pmatrix}
\ell_L^{\prime\, 2} &
\ell_L^{\prime\, 3} &
L_L^\prime
\end{pmatrix}^\intercal
\,,
\end{align}
and where $M_{q,\ell}$ are $3$-component vectors. These mass vectors can be written as 
\begin{align}
M_q=\tilde W_q\,O_q\begin{pmatrix} 0 & 0 & m_Q \end{pmatrix}^\intercal\,, \qquad
M_\ell=\tilde W_\ell\,O_\ell\begin{pmatrix} 0 & 0 & m_L \end{pmatrix}^\intercal\,,
\label{eq:M99}
\end{align}
where $m_{Q,L}$ are the vector-like fermion masses. Here, the $3\times3$ orthogonal (unitary) matrices $O_{q,\ell}$ ($\tilde{W}_{q,\ell}$) parametrize the VL-fermion mixing with the 2nd (3rd) generation, and have the form
\begin{align}\label{eq:Omatrices}
O_{q,\ell}=
\begin{pmatrix}
c_{Q,L} & 0 & s_{Q,L}\\
0 & 1 & 0\\
-s_{Q,L} & 0 & c_{Q,L}\\
\end{pmatrix}
\,,
\qquad
\tilde W_{q,\ell}=
    \begin{pmatrix}
    1 & 0\\
    0 & W_{q,\ell}
    \end{pmatrix}
    \,,
\end{align}
with $s_{Q,L}\, (c_{Q,L})$ being shorthand for the sine (cosine) of the $\theta_{Q,L}$ mixing angles, and $W_{q,\ell}$ unitary $2\times 2$ matrices. 
Once the left-handed fermion fields are redefined to diagonalize the mass terms, the interactions of the $U_1$ with the left-handed fermion fields read
\begin{align}
    \label{eq:U1int}
\mathcal{L}_U&\supset\frac{g_4}{\sqrt{2}}\, U^{\mu}_{1}   \, (\bar \Psi_q \beta_L \gamma_{\mu} \Psi_{\ell})\,,
\qquad
\beta_L = O_q \tilde{W}_q {\rm diag}(0,1,1) \tilde{W}^\dagger_\ell O_\ell^\intercal = \begin{pmatrix}
      \beta_L^{s\mu} & \beta_L^{s\tau} & \beta_L^{sL} \\
      \beta_L^{b\mu} & \beta_L^{b\tau} & \beta_L^{bL} \\
      \beta_L^{Q\mu} & \beta_L^{Q\tau} & \beta_L^{QL}
    \end{pmatrix}\,.
\end{align}
The right-handed couplings of the $U_1$ are not relevant for the current discussion and will therefore be ignored, except for their impact on the fit to the $B$-anomalies. This manifests itself in a different best-fit value for the constant $C_U = \frac{g_4^2v^2}{4m_U^2}$, which gives the overall size of the semileptonic operators, once the leptoquark has been integrated out.\\
Moreover, the field $\chi_L$ couples to the SM Higgs field and the right-handed fermions via a Yukawa interaction
\begin{equation}
\Delta\mathcal{L}_Y =  Y^\prime_- \bar{\chi}_L \psi^-_R H +  Y^\prime_+ \bar{\chi}_L \psi^+_R \tilde{H} + {\rm h.c. }  \,,
\end{equation}
with $ \tilde{H}=i\sigma_2H^\dagger$. In the mass basis before the EW symmetry breaking, it can be expressed as
\begin{equation}
\Delta\mathcal{L}_Y  \supset \nonumber
c_Q Y_- \bar{Q}_L  d_R^3 H +  c_QY_+ \bar{Q}_L  u^3_R \tilde{H}
-  s_Q Y_- \bar{q}^2_L  d_R^3 H -  s_Q  Y_+ \bar{q}^2_L  u^3_R \tilde{H}  + \rm h.c.\,,
\end{equation} where the different $Y^\prime_\pm$ and $Y_\pm$ reflect a possible mixing in the right-handed sector. Additionally, this implies $|Y_+ | \sim  y_t |V_{cb}|/ s_Q  \gg |Y_-|$, where $y_t$ is the top-quark Yukawa coupling and $V_{cb}$ denotes the element of the CKM matrix, which is why we will neglect $Y_-$ in our numerical analysis.



\section{1-loop computation}

\begin{figure}
    \centering
    \includegraphics[scale=0.7]{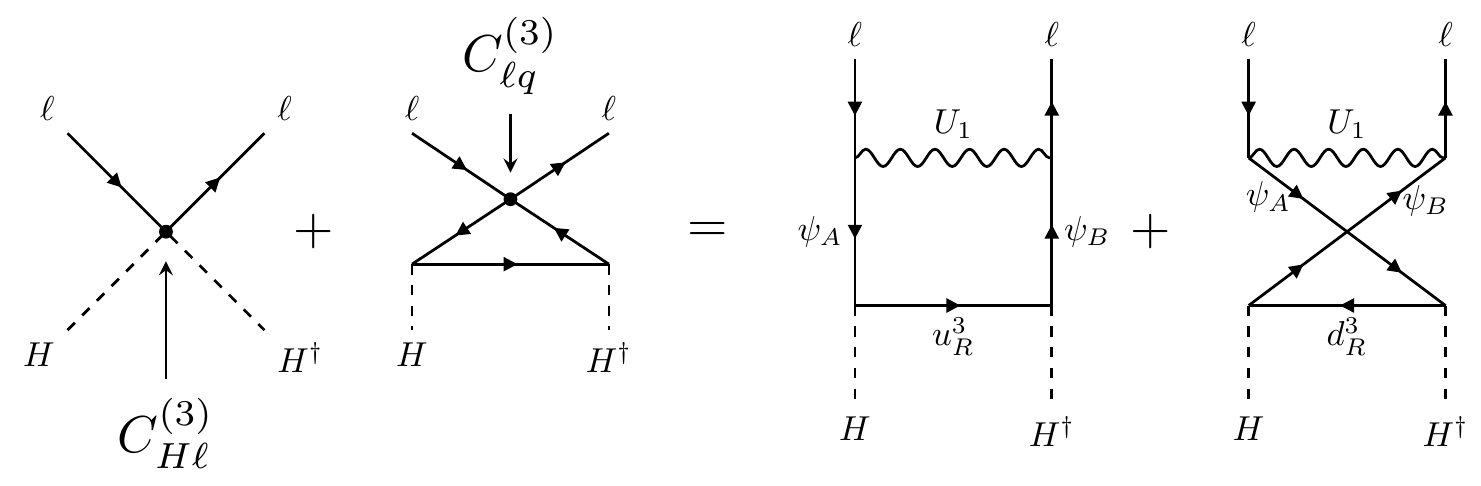}\\
    \includegraphics[scale=0.7]{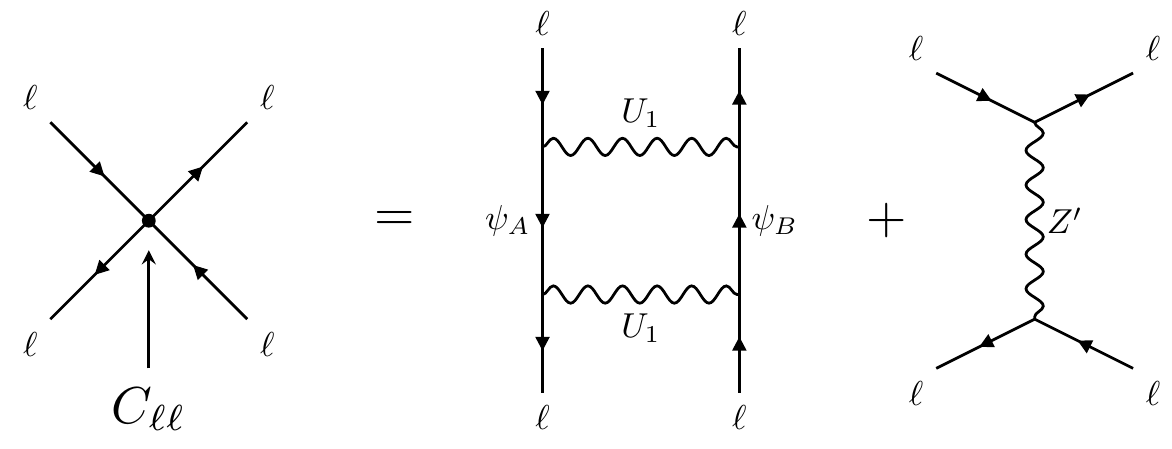}
    \caption{4321-SMEFT matching at 1-loop. $\psi_{A,B} = Q_L, q_L$ in the full theory diagrams on the right-hand side.}
    \label{fig:matching}
\end{figure}

With all the ingredients introduced so far, the 1-loop computation of the leptonic LFU ratios in the full model amounts to the computation of the 1-loop matching to the SMEFT coefficients $C_{H \ell}^{(3)}$ and $C_{\ell\ell}$. Diagrammatically, the procedure is summarized in Figure \ref{fig:matching}, and we report here only the result \footnote{Note that $[C_{H\ell}^{3}]_{ee} \approx 0$ in our framework.}:
\begin{align}
    [C_{H\ell}^{(3)}]_{\tau\tau}(\mu) &=  -\frac{1}{16\pi^2}\frac{N_{\rm c}C_U}{2} \Big[ |\beta_L^{b\tau}|^2|y_t|^2\left(1+\log\frac{\mu^2}{m_U^2}\right)
    + c_Q 2\text{Re} (\beta_L^{b\tau^*} \beta_L^{Q\tau} Y_+^* y_t) B_0(x_Q)  \\
    &+ c_Q^2 |\beta_L^{Q\tau}|^2 (|Y_+|^2 + |Y_-|^2) F(x_Q, x_Q^R) \Big]
    \\
    [C_{H\ell}^{(3)}]_{\mu\mu}(\mu) &=  -\frac{1}{16\pi^2}\frac{N_{\rm c}C_U}{2} s_L^2 \Big[ |\beta_L^{b\mu}|^2|y_t|^2\left(1+\log\frac{\mu^2}{m_U^2}\right)
    + c_Q 2\text{Re} (\beta_L^{b\mu^*} \beta_L^{Q\mu} Y_+^* y_t) B_0(x_Q)  \\
    &+ c_Q^2 |\beta_L^{Q\mu}|^2 (|Y_+|^2 + |Y_-|^2) F(x_Q, x_Q^R) \Big]
    \\
    [C_{\ell\ell}]_{\tau\mu\mu\tau}  &= 
    [C_{\ell\ell}]_{\mu\tau\tau\mu} = C_U \frac{g_4^2}{16\pi^2} s_L^2 B_{\ell\ell}^{1212} + \frac{3g_4^2 v^2}{16 m_{Z'}^2} s_\tau^2\,,
\end{align}
where $x_Q=m_Q^2/mU^2$ and $x_Q^R = m_Q^2/m_{h_U}^2$, with $h_U$ being the radial scalar-LQ excitations of $\Omega_1$ and $\Omega_3$ , and the details about the loop functions $B_0$, $F$ and $B_{\ell\ell}$ can be found in \cite{Fuentes-Martin:2020hvc,Allwicher:2021ndi}. Notice that there is also a tree-level contribution to $C_{\ell\ell}$, coming from Lepton Flavor Violating (LFV) couplings of the heavy $Z'$ boson. This is suppressed by the angle $s_\tau$, diagonalizing the lepton Yukawa couplings in the 2-3 sector, which is constrained by other observables \cite{Fuentes-Martin:2019mun}.

\section{Numerical analysis}

\begin{figure}
    \centering
    \includegraphics[width=0.48\textwidth]{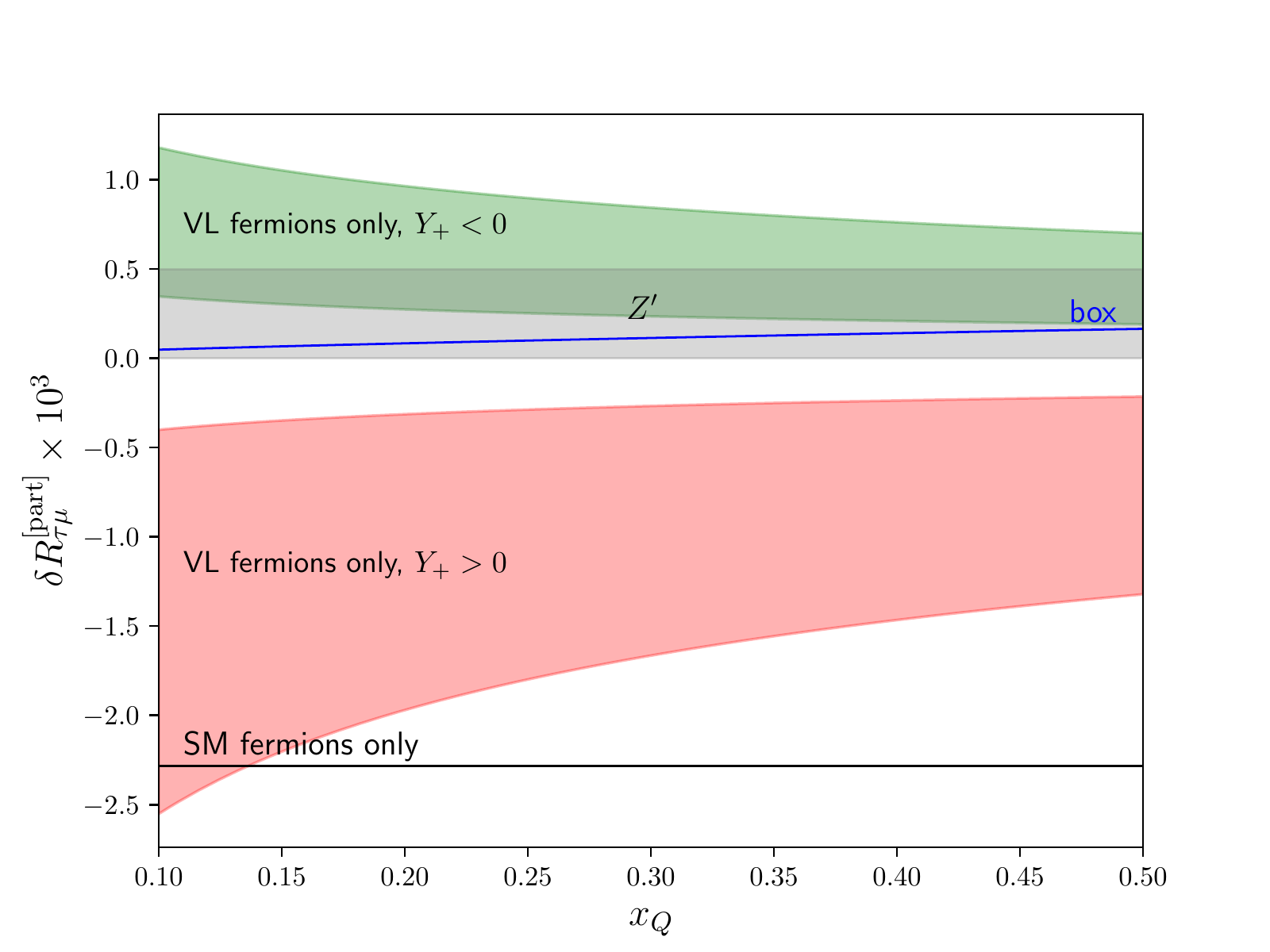}
    \includegraphics[width=0.48\textwidth]{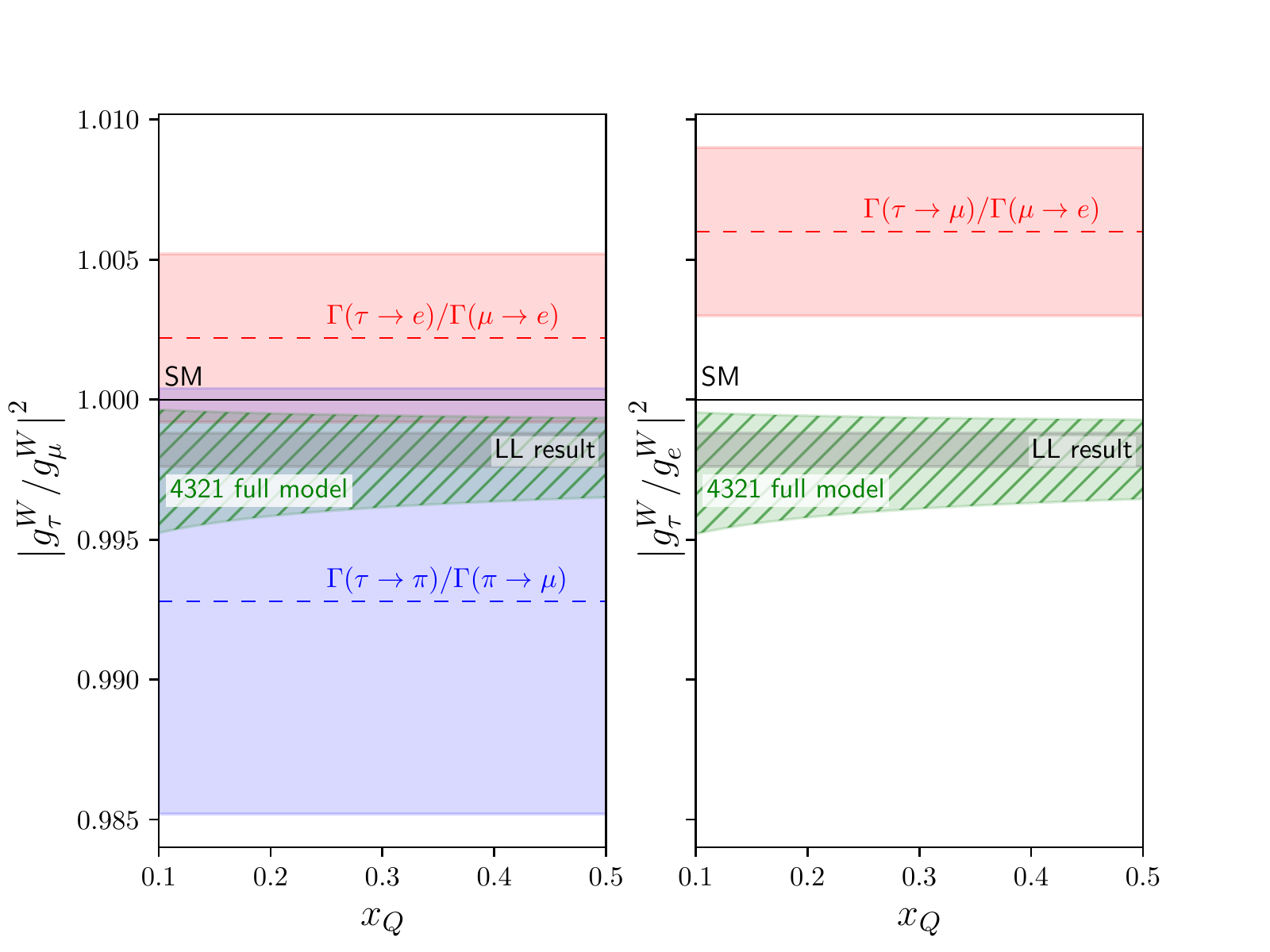}
    \caption{Numerical analysis of the modifications to leptonic decays. \emph{Left}: different contributions to $\delta R_{\tau\mu}$ as function of $x_Q = m_Q^2/m_U^2$. The black line is the contribution from the box diagrams involving SM fermions only. The red and green bands summarize all contributions involving one or more vector-like quarks for $Y_+ > 0$ and $Y_+ < 0$, respectively. The grey band is obtained varying $s_\tau$, while the blue line comes from the box diagram with two leptoquarks and four external leptons. \emph{Right}: Comparison with current experimental determination of the 4321 prediction for two LFU ratios. Both the hatched and grey bands have been obtained varying $C_U$ from 0.005 to 0.01, corresponding to the best-fit value to the $B$ anomalies in the presence or absence of right-handed currents.}
    \label{fig:Wnumerics}
\end{figure}

Using the results of the previous section, we can estimate the size of the different contributions to the leptonic LFU ratios defined in \eqref{eq:one}. 
The benchmark point for all 4321 parameters, except for $Y_+$ and $s_\tau$, is fixed by the fit to the $B$ anomalies \cite{Cornella:2021sby}. The two remaining parameters are varied in the ranges
\be
0.2 < |Y_+| < 1 \qquad 0 < s_\tau < 0.1 \,,
\ee
where the sign of $Y_+$ is also free.
Looking at Figure \ref{fig:Wnumerics} (left panel), we can see that the contributions coming from $C_{\ell\ell}$ and from the tree-level $Z'$ exchange are negligible (the latter for $0 < s_\tau \lesssim 0.07$). In this approximation, the deviations from lepton universality may directly be written in terms of modification of the $W$-couplings. Writing the $W$ Lagrangian as
\be
{\mathcal{L}}^{(\ell,W)}_{\rm eff} =-{g_\ell^W \over\sqrt{2}}~ \overline{\nu}_\ell \gamma^{\mu}  P_L \ell ~W^+_{\mu} +{\rm h.c.}\,,
\ee
we have e.g.
\be
 \left| \frac{g^{(\tau)}_e}{g^{(\mu)}_e}\right|^2  	\approx   \left| \frac{g_{\tau}^W}{g_\mu^W} \right|^2~.   
\ee
How our predictions compare to the data can be seen in Figure \ref{fig:Wnumerics} (right panel).


\noindent
Alongside with the modification of the $W$-couplings, also the $Z$-couplings to leptons receive a modification. 
A very similar computation to the one described above leads to the result
\be
\left.
\delta g^Z_{\ell_L} \right|_{Y_-=0} =  0\,,    \quad 
\left. \frac{ \delta g^Z_{\nu_\ell}}{ g^{Z,{\rm SM}}_{\nu_\ell} } \right|_{Y_-=0}
=
\left. 
\frac{ \delta g^W_{\ell}}{ g^{W,{\rm SM}}_{\ell} }\right|_{Y_-=0} \,,
\ee
where we have defined the $Z$ Lagrangian as
\be
\mathcal{L}_{\text{eff}}^{(\ell,Z)} = - \frac{g_2}{c_W} \left[ g_{\ell_L}^Z (\bar \ell \gamma^\mu P_L \ell)  + g_{\nu_\ell}^Z (\bar \nu_\ell \gamma^\mu  P_L \nu_\ell ) \right] Z_\mu\,.
\ee
\begin{figure}
	\centering
 \includegraphics[width=0.5\textwidth]{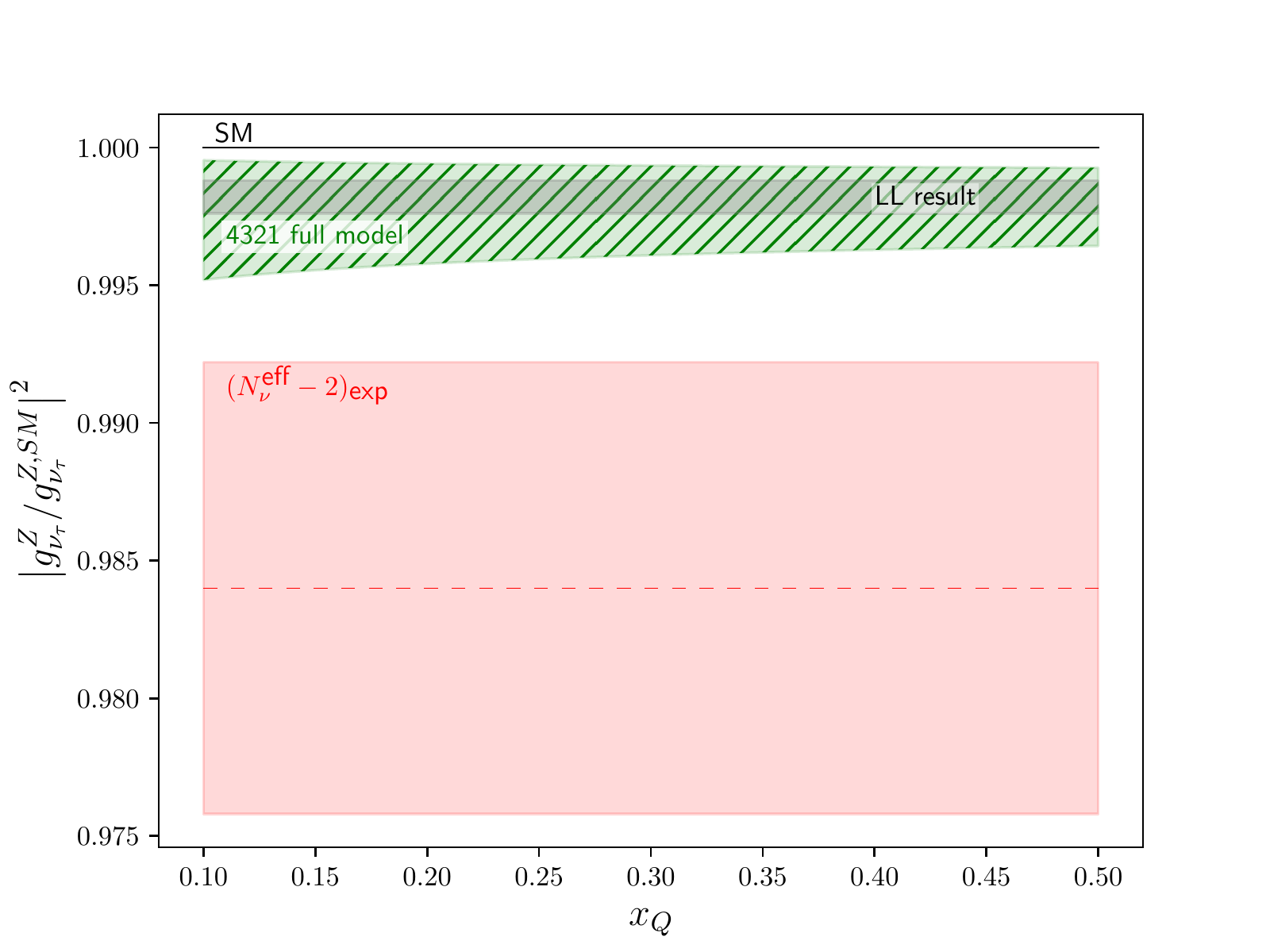}
\caption{Number of effective light neutrinos: 4321 prediction vs. experimental determination (from PDG).}
\label{fig:Znunu}
\end{figure}
\noindent This implies that the most important constraint on the model from the $Z$-pole comes from the invisible width of the $Z$, i.e. the effective number of light left-handed neutrinos. Since the only sizeable correction in our framework is the one to $\delta g^Z_{\nu_\tau}$, we have
\begin{align}
\left| \frac{  g^Z_{\nu_\tau}}{ g^{Z,\rm SM}_{\nu_\tau} }  \right|^2_{ N_\nu^{\rm eff} } = 
N_\nu^{\rm eff} - 2 \,,
\end{align}
where $N_\nu^{\rm eff}({\rm exp}) -2 =0.9840 \pm  0.0082~$. Similar to the case of the W-couplings modification, we also predict a decrease in the effective couplings of Z-boson to neutrinos, as can be seen in Figure~\ref{fig:Znunu}.

\section{Conclusion}

We have presented the first complete analysis of LFU violation in $\tau$ decays within 4321 models at next-to-leading order in the leptoquark gauge coupling. We found that the current value of the charged current $B$ anomaly always implies a decrease in the $\tau$ decay width at the few per-mil level. While being in agreement with the more general EFT expectation, the finite contributions due to the heavy vector-like states can change the effect sizeably and contribute to the agreement of 4321 models with data from leptonic decays~\cite{Cornella:2021sby}. We emphasize that the LFU tests in $\tau$ decays provide a good probe to test the model in the future, subject to a precision of $\mathcal{O}(10^{-4})$.  

\section*{Acknowledgments}

We would would like to thank Gino Isidori for help during the preparation of this talk. This work has received funding from the European Research Council (ERC) under the European Union's Horizon 2020 research and innovation programme under grant agreement 833280 (FLAY), and by the Swiss National Science Foundation (SNF) under contract 200021-175940.

\end{document}